\documentclass[12pt]{iopart}

\usepackage{subcaption}
\usepackage{graphicx}
\usepackage{epstopdf}
\usepackage{amsmath}
\usepackage{amssymb}
\usepackage{citesort}

\usepackage[dvipsnames]{xcolor}
\begin{document}

\title[Dark Sector first results at Belle II]{Dark Sector first results at Belle II}

\author{Marcello Campajola on behalf of the Belle II collaboration\footnote[1]{https://www.belle2.org/} }
\address{Dipartimento di Fisica `E. Pancini', Università di Napoli `Federico II' and INFN - Sezione di Napoli,  Via Cintia 21, I-80126 Napoli, Italy}
\ead{marcello.campajola@na.infn.it, marcello.campajola@unina.it}
\vspace{10pt}
\begin{indented}
\item[]January 2021
\end{indented}

\begin{abstract}

Understanding the nature of dark matter is one of the most exciting challenges in fundamental physics nowadays, requiring the synergy of different search techniques, as well as theoretical inputs. An interesting opportunity for the investigation of dark matter is the one offered by the \textit{B}-factories. The Belle~II experiment, located at the interaction point of the SuperKEKB asymmetric energy  $e^+e^-$ collider, is a new generation \textit{B}-factory experiment operating at the Japanese KEK laboratory.
With a machine design luminosity of $6\times10^{35}\,\rm{cm}^{-2}s^{-1}$,  Belle II  aims to record 50 ab$^{-1}$ of data within the next decade.
Thanks to this large data-sample and by using dedicated triggers, Belle~II is expected to explore dark sector candidates with  unprecedented sensitivity in the mass range up to 10 GeV$/c^2$.
During 2018, the experiment concluded a commissioning run, collecting a data-sample corresponding to an integrated luminosity of about 0.5 fb$^{-1}$, while main operations started on March 2019 with an almost complete detector. So far the experiment collected an integrated luminosity of $\sim90\, \rm{fb}^{-1}$. 
With these data-sets Belle~II has already shown the possibility to search for a large variety of dark sector candidates in the GeV mass range.
This paper  reviews the status of the dark sector searches performed at the Belle~II experiment, with a focus on the first obtained results and the discovery potential with the  data-set available in the short term.
\end{abstract}

%
\vspace{2pc}
\submitto{\PS}
%
%
%

\section{Introduction}
A large number of astrophysical and cosmological observations, from galaxies to larger-scale structures
in the Universe, suggests the existence of \textit{Dark Matter} (DM), i.e., a kind of matter whose constituents do not interact through strong or electromagnetic forces. It has been estimated that DM accounts for $\sim$27\% of the total energy density of the Universe and for about 85\% of its matter density \cite{refId0}.
While important clues about DM have been inferred in recent years, the fundamental nature of the DM is still a puzzling mystery, providing one of the most important open problems in fundamental physics nowadays. 
DM cannot be incorporated within the Standard Model (SM) of particles, and its understanding requires, in most of the theoretical scenarios, the introduction of new degrees of freedom.  

Although dark matter has traditionally been associated with heavy candidates, Weakly Interacting Massive Particles (WIMPs) for example, a class of lighter particles remains an interesting alternative.
In recent times many theoretical models have been proposed to account for light dark matter, from minimal scenarios including a single dark matter particle to the case of an entire structure decoupled from SM, a so-called \textit{Hidden} or \textit{Dark Sector}.
The dark sector may contain new light dark matter states well below the weak-scale, that interact feebly with ordinary matter, so that they could easily have escaped past experimental searches.
The connection between SM and the dark sector is usually made through a so-called `portal', i.e., a particle which possesses both Standard Model and dark sector quantum numbers, and interacts with SM particles either directly 
or indirectly through loop diagrams or mixing. 
Depending on the properties of such a particle, several scenarios are possible, and a few categories of models are available (see e.g. \cite{1483008,Batell:2009di} for a review): the so-called ``vector" portal which introduces a new vector gauge mediator, a dark photon $A^\prime$ or a $Z^\prime$ as an example; the ``neutrino" portal where a sterile neutrino interacts with SM neutrino with a Yukawa coupling term; the ``scalar" portal which gives rise to a Higgs-like mediator coupling to the SM Higgs boson; or the ``pseudo-scalar" portal where interaction is mediated by axions (or axion-like particles), for example.

Other than  predicting the observed DM abundance, the introduction of a dark sector could explain other features: it could be invoked to explain the antimatter excess in the cosmic rays \cite{Adriani:2008zr,FermiLAT:2011ab,PhysRevLett.110.141102} as well as some SM anomalies, like the discrepancy between the theoretical and experimental values for the muon anomalous magnetic moment \cite{Bennett:2006fi}.

In recent times, dark sector searches at collider experiments have aroused a lot of interest.
To date, most of the searches have been performed at the high-energy frontier experiments, like those at LEP, LHC and Tevatron \cite{Boveia:2018yeb}. Such experiments are ideally suited to probe weak-scale dark matter candidates. 
In contrast, low-energy $e^+e^-$ colliders like \textit{B}-factories, as they operate at a much lower center-of-mass energy, have higher sensitivity to probe light dark sector candidates \cite{Borodatchenkova:2005ct,Fayet:2007ua,Batell:2009yf,Essig:2009nc,Essig:2013vha,Reece:2009un,Yin:2016hbi}. The Belle~II experiment at the Japanese KEK laboratory, is a new generation \textit{B}-factory experiment, which is expected to  explore dark sector physics with unprecedented precision in a mass range up to $\sim$ 10 GeV$/c^2$. 
It must be mentioned that the landscape  of light dark sector searches includes also many other experimental possibilities. As an example, fixed target experiments are very well suited for the detection of light dark sector particles, produced either in beam interaction with the target nuclei or in the decay of secondary particles, like the experiments APEX, HPS or NA64. Even neutrino experiments, such as MiniBooNE, have good sensitivities to the DS. On the other hand, light dark sector searches are performed even at high energy collider experiments, despite their better sensitivity for higher masses. Furthermore, GeV and sub-GeV DM candidates can be explored even in direct detection experiments, although  hampered by the small energy depositions. For an exhausting review of past and planned experimental efforts for light dark sector searches see, for example,  \cite{Alexander:2016aln,Agrawal:2021dbo,Battaglieri:2017aum}.

This paper aims to give an overview of the dark sector searches performed or capable of being performed at the Belle~II experiment. 
Here, we  will focus on the first results obtained with the early data-sets collected by Belle~II so far and on the discovery potential of the  data-set which will be available in the short term.

\section{Status of the Belle II experiment}
Belle II is a full upgrade of the Belle experiment operating at the KEK laboratory (Japan). It is located at the interaction region of the SuperKEKB machine \cite{AKAI2018188}, an $e^+e^-$ energy-asymmetric collider that operates  at a centre-of-mass system (CMS) energy of 10.58 GeV, which corresponds to the $\Upsilon(4S)$ resonance mass. As this latter decays mostly into a pair of \textit{B} mesons, Belle II and SuperKEKB are called a \textit{B}--factory experiment.
Specifically, they are a second generation \textit{B}-factory, where both the detector and the machine have undergone a renewal compared to the predecessor experiment (Belle at KEKB).
The higher beam currents and the smaller interaction region with the application of the large crossing angle nano-beams scheme \cite{Bona:2007qt} will allow SuperKEKB to provide a significant increase in the instantaneous luminosity with respect to KEKB. Values of $6\times10^{35}\,\rm{cm}^{-2}\rm{s}^{-1}$ are expected soon.

Belle II inherits the design of the Belle detector \cite{ABASHIAN2002117} with major improvements in all of the sub-systems. 
It consists of different layers of sub-detectors arranged 
in a cylindrical geometry around the interaction region. 
The innermost sub-detector is the silicon vertex detectors (VXD), surrounded by a large helium-based small-cell drift chamber (CDC). A particle identification
system follows, based on an imaging Time-Of-Propagation (TOP) detector in the barrel region and an Aerogel Ring Imaging Cherenkov (ARICH) detector in the forward endcap region. A CsI(Ti) segmented electromagnetic calorimeter (ECL) is used to measure the energy of photons and electrons. The outermost detector, called KLM, is made of scintillator strips and resistive plate chambers and serves for $K_L$ and muon reconstruction. 
All of the sub-detectors with the exception of the KLM are surrounded by a superconducting magnet which produces a  solenoid field of 1.5 T, in the direction parallel to the detector axis.
The detector is described in more detail in Ref.~\cite{10.1093/ptep/ptz106}. 

Belle II and SuperKEKB are expected to achieve an integrated luminosity of 50~ab$^{-1}$ by $\sim$ 2030, with a large portion of the data collected at the nominal collision energy ($\sqrt{s} = 10.58\,$GeV). 
While the primary purpose of Belle II is to study the properties of \textit{B}-mesons, the experiment is ideally suited for a wide range of new-physics searches, including dark sector physics. Indeed, because of the simple and clean initial state, Belle~II is very sensitive to final states with invisible particles, thus having excellent potential for  dark sector searches. In addition, a major upgrade compared to previous \textit{B}--factories experiments comes from new triggers specifically designed to have high efficiency for events with a low number of particles in the final state: most relevant  is the newly designed single-photon trigger.  
The expected large data-set and the clean $e^+e^-$ environment, combined with a trigger specifically designed for dark sector signature, will enable Belle II to explore dark sector physics with unprecedented precision in a mass range up to $\sim$ 10 GeV$/c^2$. 
The dark sector physics search  program at Belle~II is broad and foresees to significantly extend the range of parameters covered by previous low-energy $e^+e^-$ experiments, namely BaBar, Belle, BESIII and KLOE.

The experiment performed the first data-taking runs for physics analysis during 2018. During this period, the experiment worked  with a partially installed vertex detector at a reduced instantaneous luminosity for commissioning purposes, and an integrated luminosity of approximately 0.5 fb$^{-1}$ was collected \cite{Abudin_n_2020}. In early 2019, Belle II started the main operations with a near-complete detector and  collected approximately 90 fb$^{-1}$ of data so far.

This paper reviews the first dark sector  searches performed by Belle~II with the 2018 data, namely the searches for an ALP decaying into two photons \cite{BelleII:2020fag} and for an invisibly decaying $Z^\prime$ \cite{Adachi:2019otg}, and the forthcoming searches in preparation for the new data collected so far or  available in the short term. The analysis selections for each analysis are described and results or sensitivities in case results were not yet available, are shown. The analysis optimizations rely on an intensive use of Monte Carlo simulations, where the interaction of particles with the detector and  reconstruction  efficiencies are determined by means of \textsc{Geant4} \cite{Agostinelli:2002hh}, while to  reconstruct and analyse events  the  Belle~II  Analysis Software  Framework  is used \cite{Kuhr:2018lps}.

\section{Axion-like Particles}
With the expression Axion-like particles (ALPs) one refers to a class of hypothetical pseudo-scalar ($J^P=0^-)$ particles that occur in several extensions of the SM \cite{jaeckel}. Differently from axions, introduced to preserve some of the QCD properties \cite{Peccei}, the mass and the coupling of ALPs are independent. 
ALPs are often invoked as a portal between SM and yet undiscovered DM particles \cite{Nomura:2008ru} or sometimes considered as cold dark matter  them-self \cite{Arias:2012az}, solving several astrophysical anomalies, like the x-ray line at $\sim$3.5 keV \cite{Cicoli:2014bfa,Jaeckel:2014qea}.

Several models have already been excluded from collider or beam-dump experiments. However, cases where they predominantly couple to $\gamma\gamma$, $\gamma Z^0$, and $Z^0Z^0$ are experimentally much less constrained than those that couple to gluons or fermions. 
Among the former, the most accessible at a low energy $e^+e^-$ collider is the case where ALPs couple to photons. In particular, there are two different production processes of interest at a low energy $e^+e^-$ collider \cite{torben}: the ALP--strahlung process $e^+e^-\to\gamma^*\to\gamma\,a$  and the photon-photon fusion $e^+e^-\to e^+e^- a$. 

\subsection{Two-photon decay}
\begin{figure}
\begin{subfigure}{.45\textwidth}
  \centering
    \includegraphics[width=0.99\textwidth]{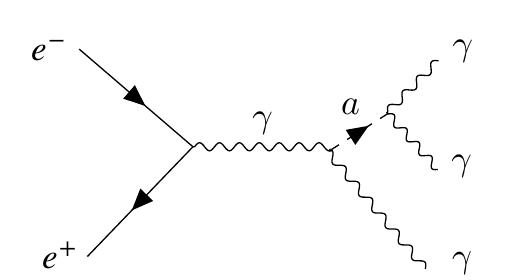}
    \caption{}
  \label{fig:ALP_feynman}
\end{subfigure}
\quad
\begin{subfigure}{.45\textwidth}
  \centering
    \includegraphics[width=0.99\textwidth]{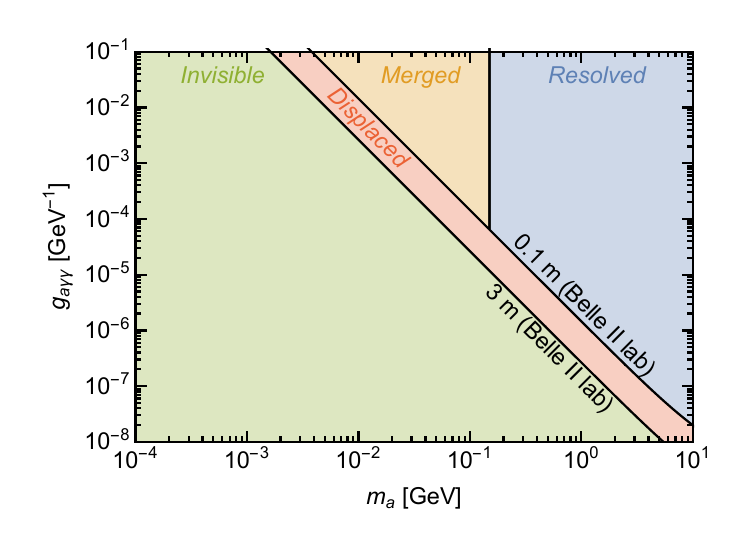}
    \caption{}
  \label{fig:ALP_topology}
\end{subfigure}
\caption{ (a) Feynman diagram for the ALP-strahlung process with the following ALP decay to a photon pair. 
(b) Signatures of the ALP decay into two photons  with the Belle II detector, as a function of the ALP mass $m_{a}$ and of the coupling constant  $g_{a\gamma\gamma}$ \cite{torben}. }
\label{fig:ALP}
\end{figure}

Among the first results obtained by Belle II is the search for ALPs produced by the ALP--strahlung process (see Fig.~\ref{fig:ALP_feynman}), with the following decay into two photons. 
Unless $m_a$ is close to $\sqrt{s}$, the cross-section for the ALP production via the photon-fusion process is larger that the ALP-strahlung one. Nevertheless, this latter is much harder to be detected from an experimental point of view: ALPs carry little energy, thus decaying into relatively soft photons, which result in large QED backgrounds \cite{torben}.
For this search, it was assumed a predominant coupling to photons with strength $g_{a\gamma\gamma}$, and a negligible coupling strength $g_{a\gamma Z}$ to a photon and a $Z^0$ boson, so that $BR(a\to\gamma\gamma)=100\%$.

The experimental signature of the two-photon decay  depends on  $m_a$ and  $g_{a\gamma\gamma}$, as illustrated in Figure~\ref{fig:ALP_topology}. They affect both the ALP decay length  and the opening angle between the photons \cite{torben}. 
When the ALP mass is high, a large opening-angle of the decay photons is expected, leading to a final state with three detectable photons, denoted as   \textit{Resolved}. For small ALP masses, the higher boost leads to a decreased
opening-angle between the ALPS decay photons that cannot be resolved  for $m_a\lesssim200\,\rm{MeV/c^2}$, due to the limited spatial resolution of the electromagnetic calorimeter; the phase space corresponding to such a case is referred to  as \textit{Merged}.
\textit{Invisible} denotes the case where the ALP is long-lived, decaying outside of the detector volume. Here, only one photon can be detected. 
A further possibility, denoted by \textit{Displaced}, represents the case when the ALP decays far from the interaction point but still within the Belle~II detector.

Belle  II  searched for an ALP decay into two photons produced with the ALP--strahlung process ($e^+e^-\to\gamma \,a, \,a\to\gamma\gamma$) over a mass range 0.2 $<$ $m_a$ $<$ 9.7 GeV/c$^2$ in the \textit{resolved} three-photon final state \cite{BelleII:2020fag}. The entire data-set collected during 2018 was used for this analysis. A subset of data was used to validate the selection and then discarded; the remaining part used for this search corresponds to 0.445 pb$^{-1}$ of data.
The final state consists of three photons, with two of them peaking at the ALP mass, while the recoiling one being a mono-energetic photon in the CM frame of energy $E^{\rm{CM}}_{\rm{rec}_\gamma}=(s-m_a^2)/2\sqrt{s}$.
The ALPs signature would be a narrow peak in the squared mass distribution of the recoiling system against the mono-energetic photon  $M^2_{\rm{rec}}=s-2\sqrt{s}E^{\rm{CM}}_{\rm{rec}_\gamma}$, or in the squared-invariant-mass distribution of the two decay photons $M^2_{\gamma\gamma}$.

The analysis selection and the statistical interpretation procedures have been optimized on Monte Carlo simulation prior to examining data.
The dominant SM background comes from the QED process $e^+e^-\to\gamma\gamma\gamma$. A smaller background source  arise especially for small ALP masses from $e^+e^-\to\gamma\gamma$  with an additional photon coming from beam-induced backgrounds, or in the case where one of the two photons produces  an $e^+e^-$ pair outside of the tracking detectors, resulting in both electron and positron identified as photons. A further contribution comes from $e^+e^-\to e^+e^-\gamma$ due to tracking inefficiencies.
Such background processes have been generated with \textsc{Babayaga@nlo} \cite{Balossini:2008xr}. Further background sources are due to the processes $e^+e^-\to P\gamma(\gamma)$ with $P=\pi^0,\eta,\eta^\prime$, which have been generated by means of \textsc{Phokara9} \cite{Czyz:2017veo}. Signal events  have been generated with the \textsc{MadGraph5} generator \cite{Alwall:2014hca}, by including also the   initial-state-radiation effect in the kinematics of the events \cite{Li:2018qnh}. Samples with different  mass hypothesis $m_a$ have been generated over the full search range with a step of approximately the mass resolution. 
The analysis selection requires three-photon candidates with a combined invariant mass close to the center-of-mass energy. The photon candidate polar angle must be within a restricted ECL barrel region ($37.3^\circ<\theta_\gamma<123.7^\circ$) in order to avoid regions close to ECL gaps. Furthermore, a good energy resolution and a low background is expected in such a region. 
Only photons reconstructed in time with each other are allowed and events with track coming from the interaction region are rejected. Furthermore a shower-shape multivariate based on multiple Zernike moments \cite{ZERNIKE1934689} on the most isolated photon candidate is applied. 

After  the  final  selection,  the expected background distributions are dominated by $e^+e^-\to\gamma\gamma\gamma$, with a small contribution from  $e^+e^-\to e^+e^-\gamma$.
A mass scan technique  has been adopted to search for peaks as a function of $m_a$ by performing a series of independent binned maximum-likelihood fits. 
The resolution of $M_{\gamma\gamma}$ worsens with increasing $m_a$, while that of $M^2_{\rm{rec}_\gamma}$ improves. Thus, fits have been performed on the $M_{\gamma\gamma}$ invariant mass  in the ALP low-mass region ($m_a < 6.85\, \rm{GeV/c^2}$), while  on the recoil spectrum $M^2_{\rm{rec}_\gamma}$ for the higher part ($m_a > 6.85\, \rm{GeV/c^2}$).

\begin{figure}
\begin{subfigure}{.5\textwidth}
  \centering
    \includegraphics[width=0.88\textwidth]{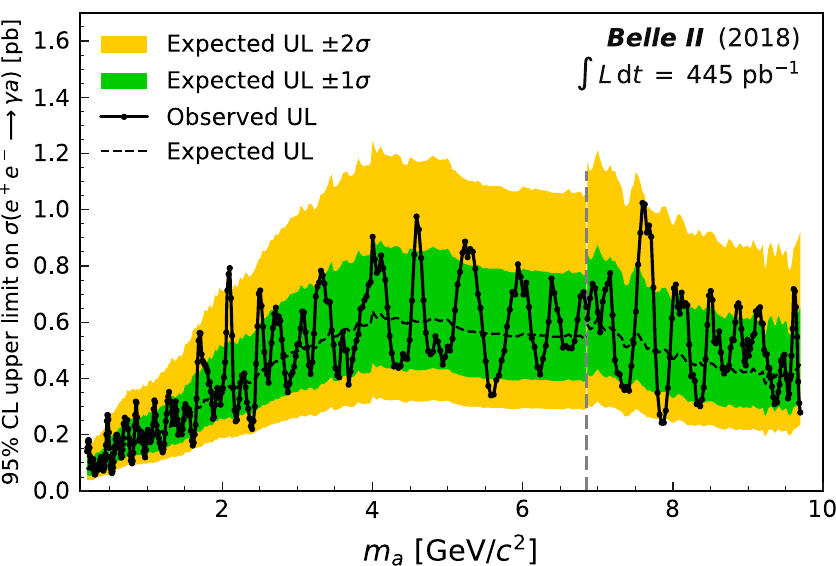}
    \caption{}
  \label{fig:ALP_UL_CS}
\end{subfigure}%
\begin{subfigure}{.5\textwidth}
  \centering
    \includegraphics[width=0.950\textwidth]{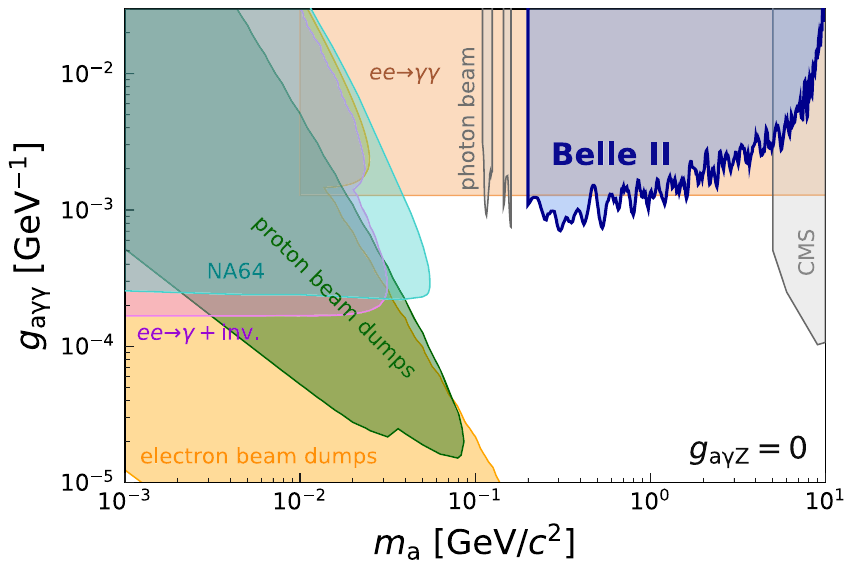}
    \caption{}
  \label{fig:ALP_UL_coupling}
\end{subfigure}
\caption{ Results for the  $e^+e^-\to\gamma \,a, \,a\to\gamma\gamma$   search performed with the 2018 data-set (455 pb$^{-1}$) \cite{BelleII:2020fag}. (a)
95\% C.L. expected and observed upper limits  on the
ALP cross section $\sigma_a$. The vertical dashed line at 6.85 GeV/c$^2$ indicate the change in the in the invariant-mass determination method.
(b) 95\% C.L. observed upper limit to the  $g_{a\gamma\gamma}$ coupling constant from the Belle~II analysis. Constraints from 
$e^+e^-\to\gamma+\rm{invisible}$ \cite{torben}, $e^+e^-\to\gamma\gamma$ \cite{Knapen:2016moh}, beam dump experiments \cite{Banerjee:2020fue,Dobrich:2015jyk}, heavy-ion collisions \cite{Chudasama:2019jmm} and photon-beam experiments \cite{Aloni:2019ruo}  are also shown.}
\label{fig:ALP}
\end{figure}

No significant excesses of events consistent with an ALP signal have been observed in data.
Thus the 95\% confidence level (C.L.) upper limits on the cross section as a function of $m_{a}$ using a one-sided frequentist profile-likelihood method have been computed (Figure~\ref{fig:ALP_UL_CS}), and then converted in terms of the $g_{a\gamma\gamma}$ coupling constant.
These latter limits are shown in Figure~\ref{fig:ALP_UL_coupling}  together with the existing constraints from previous experiments.
Belle~II  limits are more stringent than the previous ones for $0.2 < m_a < 1$ $\rm{GeV/c}^2$.
For ALP masses approaching the CM energy $\sqrt{s}$, the cross section vanishes \cite{torben}, limiting the sensitivity to the $g_{a\gamma\gamma}$ coupling constant  in the high mass region. 
In a future update of the analysis with the full data-set, Belle~II is expected to improve the limits to the $g_{a\gamma\gamma}$ coupling constant by more than one order of magnitude \cite{torben}.

\subsection{Invisible  decay}
The Belle~II program for ALPs searches can be extended by taking into account also the case of ALPs coupled to DM particles $\chi$ \cite{Nomura:2008ru,torben}. 
In such a scenario, provided the DM  is sufficiently light ($m_\chi<1/2\,m_a$), the ALP has an invisible decay mode $a\to\chi\bar{\chi}$.
Thus one can look at the ALP invisible decay produced with the ALP--strahlung process ($e^+e^-\to\gamma \,a, \,a\to \rm{invisible}$). 
This  signature is of particular interest since, in the case of DM kinematical accessible, one can assume the invisible decay mode to be dominant.
In such a case, the final state consists of a single mono-energetic photon  in the CM frame of energy $E^{\rm{CM}}_{\rm{rec}_\gamma}=(s-m_a^2)/2\sqrt{s}$.

\begin{figure}
  \centering
    \includegraphics[width=0.45\textwidth]{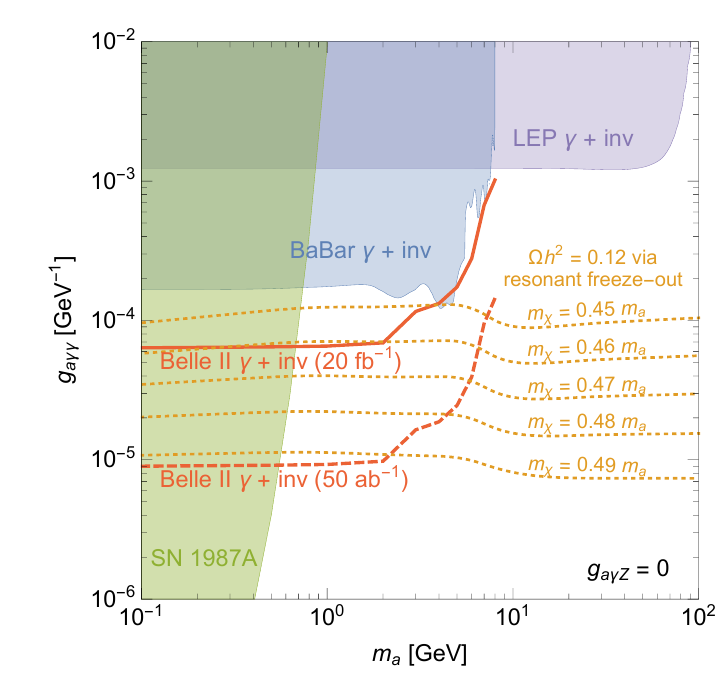}
    \caption{90\% C.L. expected upper limits to the $g_{a\gamma\gamma}$ coupling constant via the process $e^+e^-\to\gamma \,a, \,a\to\chi\bar{\chi}$ \cite{torben}. The yellow dashed lines represents the parameter region able to explain the DM relic abundance via resonant DM annihilation into photons \cite{Nomura:2008ru,torben}. This latter process is efficient only for $m_\chi$ slightly smaller than $1/2\,m_a$.}
  \label{fig:ALP_DM}
\end{figure}

This search has the same experimental signature of the dark photon decays into DM. The main difference is in the angular distribution of the recoil photon, which at small angles is less peaked than the ISR photon in the radiative dark photon process. 
Single-photon searches at Belle~II are currently under development.
Using the the same selection criteria as for the invisible dark photon search (described below), the expected upper limits to $g_{a\gamma\gamma}$  via the process $e^+e^-\to\gamma \,a, \,a\to\chi\bar{\chi}$ have been computed for the case of 20 fb$^{-1}$ and 50 ab$^{-1}$, as shown in Figure~\ref{fig:ALP_DM}. For ALP masses approaching to $\sqrt{s}$, the cross section vanishes \cite{torben}, limiting the sensitivity to the $g_{a\gamma\gamma}$ coupling constant. Furthermore, the sensitivity to the high mass parameter region is  limited by the energy threshold for the single-photon trigger, here considered at 1.8 GeV (more details on the single-photon trigger are given in Section~\ref{sec:SP}). Results on data are expected in a short-term period.

\section{$Z^\prime$ ($L_\mu-L_\tau$ model)}
New physics can  elegantly be introduced into the Standard Model by gauging some of its accidental global symmetries. This is the case of the $L_\mu-L_\tau$ model, which gauges the leptonic muon and tau number difference \cite{He:1991qd}. As a result, a new massive gauge boson $Z^\prime$ is introduced,  which  couples only to the 2$^{\rm{nd}}$ and 3$^{\rm{rd}}$ generation leptons with a new coupling constant indicated with $g^\prime$.  Such a boson does not couple with $e$ and $\nu_e$ and thereby may have escaped existing searches.
This model is of particular interest, being able to explain various phenomena not accounted for within the SM, i.e., the problem of DM abundance by providing a way to balance the annihilation rate to sterile neutrinos in the early universe \cite{Shuve:2014doa,Altmannshofer:2016jzy}, the $(g-2)_\mu$ anomaly \cite{Baek:2001kca,Ma:2001md,Altmannshofer:2016brv} and some  anomalies in the $b\to s\mu^+\mu^-$  decays \cite{Altmannshofer:2016jzy,Crivellin:2015mga,Baek:2017sew} reported by the LHCb experiment \cite{Aaij:2013qta}.

At an $e^+e^-$ collider the $Z^\prime$ would be produced in processes such as $e^+e^-\to\mu^+\mu^-Z^\prime$ or $e^+e^-\to\tau^+\tau^-Z^\prime$,  thus being radiated from one of the final state muons or taus and then eventually decaying either visibly into a muon or tau pair, or invisibly to neutrinos or dark matter \cite{Zhang:2020fiu}. 
Related searches have been performed by the BABAR and CMS experiments for a $Z^\prime$ produced with a muon pair and decaying to muons \cite{PhysRevD.94.011102,Sirunyan:2018nnz}.

\subsection{Invisible Decay}
\begin{figure}
  \centering
    \includegraphics[width=0.48\textwidth]{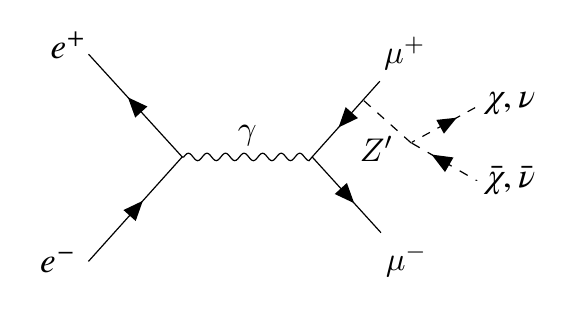}
    \caption{Feynman diagram for a $Z^\prime$ production with a muon pair and  the  following decay to invisible.}
  \label{fig:Zprime_feynman}
\end{figure}

Belle~II investigated  the invisible decay topology \cite{Adachi:2019otg} via the process $e^+e^-\to\mu^+\mu^-Z^\prime,\,Z^\prime\to \rm{invisible}$   where the $Z^\prime$ is radiated from one of the two muons (as shown in Figure~\ref{fig:Zprime_feynman}) for the first time.
The final state consists of two muon tracks with opposite charge coming from the interaction point plus missing energy.
According to the $L_\mu-L_\tau$ model, the $Z^\prime$ decay branching ratio  depends on the boson mass $m_{Z^\prime}$ as follow: if $m_{Z^\prime}<2m_\mu$, the only possibility is the invisible decay to neutrinos and $BR(Z^\prime\to \rm{invisible}) = 1$. If $m_{Z^\prime}>2m_\tau$ both visible decays in muons and taus and the invisibile decay are allowed, thus $BR(Z^\prime\to \rm{invisible}) \sim 1/3$ \cite{Curtin:2014cca}. 
However, in case of kinematically accessible DM candidates and if they are charged under  $L_\mu-L_\tau$, due to the expected much stronger coupling with respect to SM particles, the invisible decay branching ratio is enhanced to 1. 
Invisible decay topology is then a probe for the existence of  new invisible particles.

This search has been performed by Belle II using the data collected  during the 2018 pilot run. Due to the low-multiplicity trigger configuration usable for this search, only 0.276 fb$^{-1}$ were available. 
The final state is reconstructed by knowing the initial $e^+e^-$ state energy and the properties of the muon tracks in the CM frame with high accuracy. A recoil mass against the muon pair is defined as $M^2_{\rm{rec}}=s+M^2_{\mu\mu}-2\sqrt{s} E_{\mu\mu}^{\rm{CM}}$, where $M^2_{\mu\mu}$ is the squared invariant mass of the muon pair and $E^{\rm{CM}}_{\mu\mu}$ is the sum of the muon energies in the CM frame.
In the case of signal events, `recoil' quantities coincide with the $Z^\prime$ features; thus a peak in the $M^2_{\rm{rec}}$ distribution corresponding to the $Z^\prime$ mass is expected.

The backgrounds are SM final states with two tracks identified as muons and missing energy due to undetected particles. 
The primary sources come from QED processes such as $e^+e^-\to\mu^+\mu^-(\gamma)$ with one or more photons not detected due to inefficiencies or being out of the detector acceptance,  $e^+e^-\to\tau^+\tau^-(\gamma)$ with $\tau \rightarrow \mu\bar{\nu_\mu}\nu_\tau$ or $\tau \rightarrow \pi\nu_\tau$ (due to pions misidentification in muons) where missing momentum comes from neutrinos from both $\tau$ decays, and  $e^+e^-\to e^+e^-\mu^+\mu^-$ with electron and positron usually are outside the detector acceptance. 
The selection has been optimized using simulated samples before to examining data. Muon and tau pair background sources have been simulated with the \textsc{KKMC} generator \cite{Jadach:1999vf}, while four leptons with \text{AAFH} \cite{Berends:1984gf}.
For signal events, different $M_{Z^\prime}$ mass hypothesis  with a regular step size of 0.5~GeV/c$^2$ have been generated with \textsc{MadGraph5}.

\begin{figure}
\begin{subfigure}{.5\textwidth}
  \centering
    \includegraphics[width=0.950\textwidth]{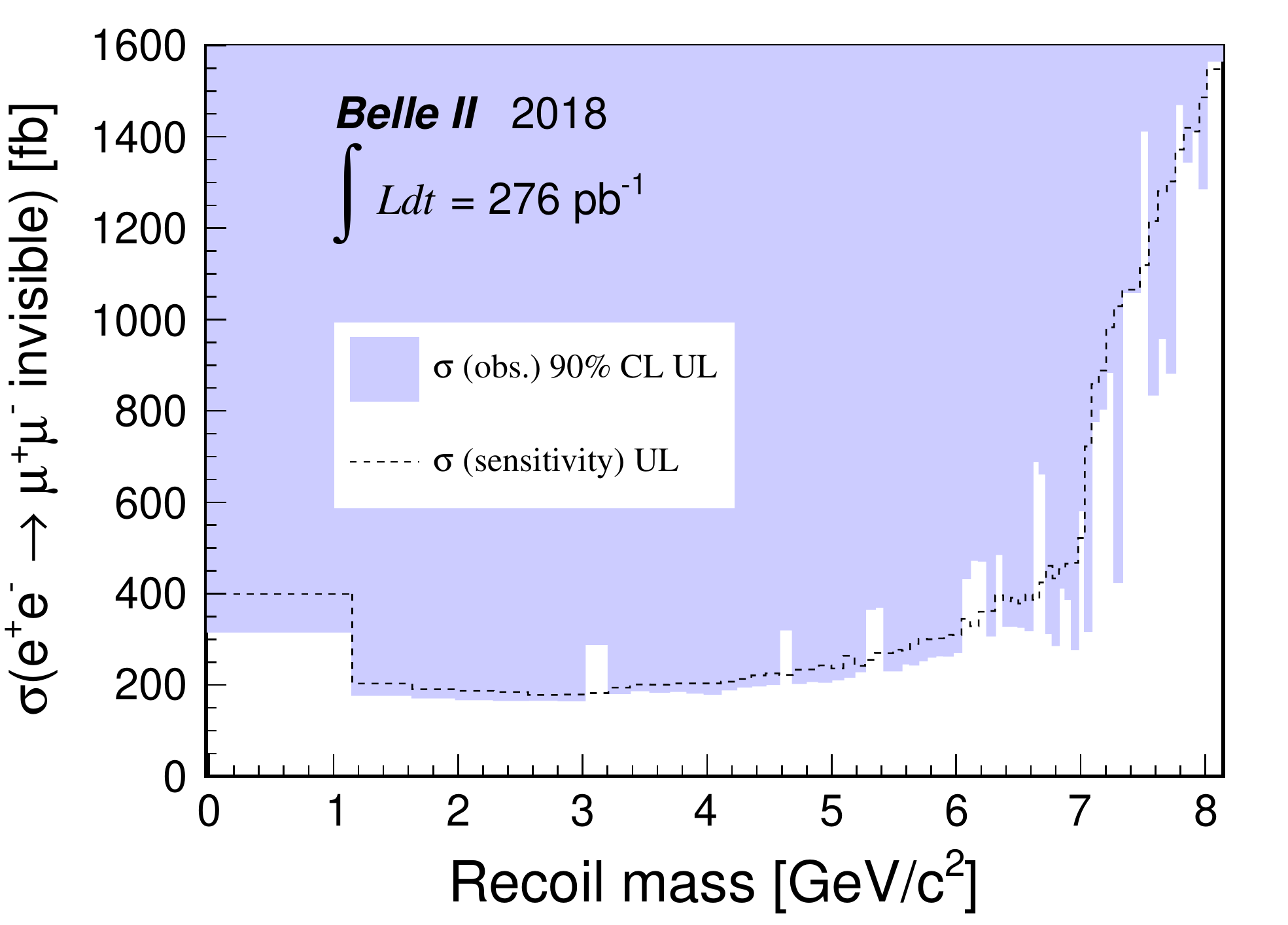}
    \caption{}
  \label{fig:Zprime_rec}
\end{subfigure}%
\begin{subfigure}{.5\textwidth}
  \centering
    \includegraphics[width=0.950\textwidth]{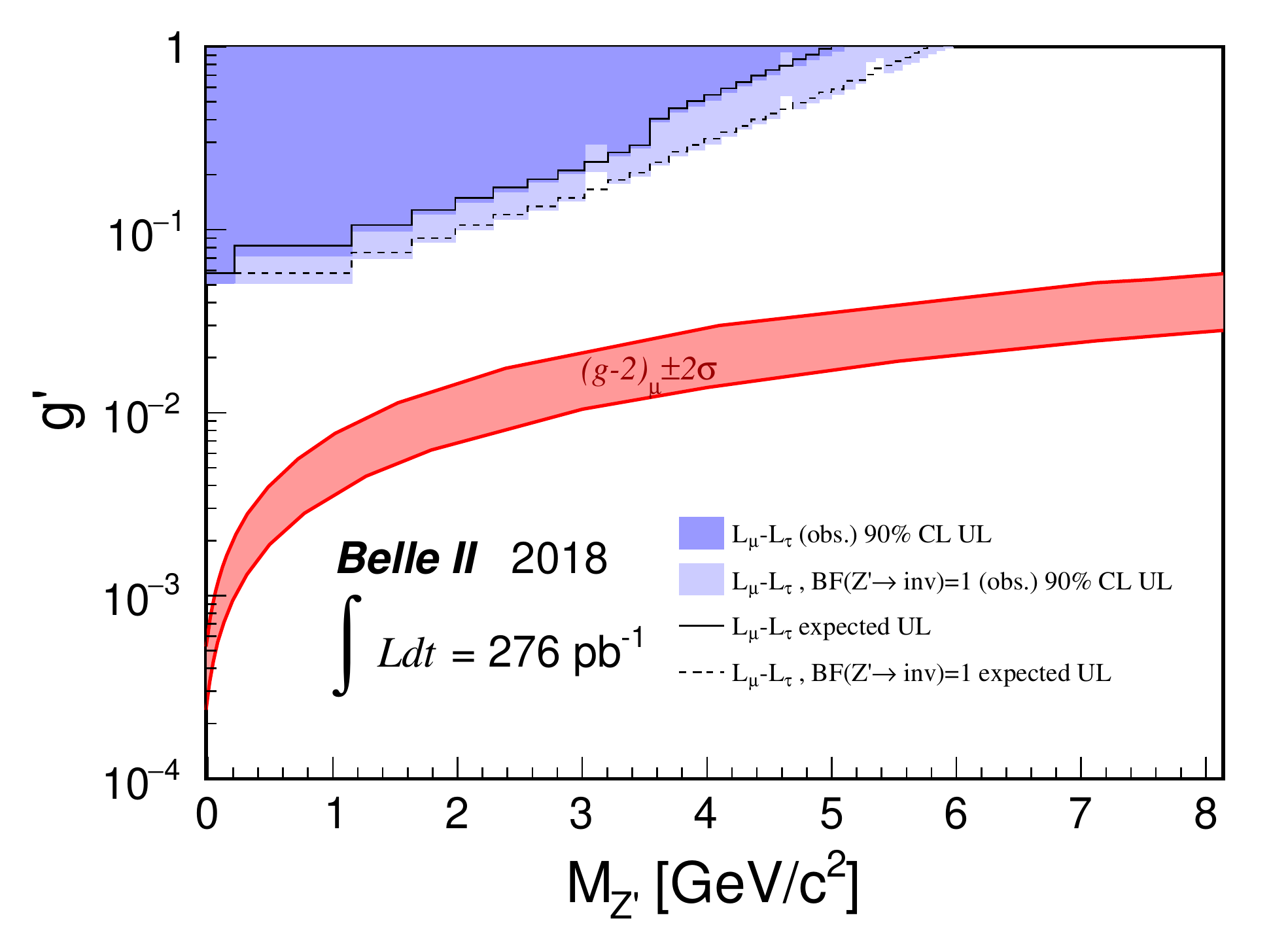}
    \caption{}
  \label{fig:Zprime_up}
\end{subfigure}
\caption{ Results for the  $e^+e^-\to\mu^+\mu^-Z^\prime,\,Z^\prime\to \rm{invisible}$   search performed with the 2018 data-set ($\int L\,dt=276\, \rm{pb}^{-1}$) \cite{Adachi:2019otg}. (a)
90\% C.L. upper limits on the cross section. The dashed line represents the expected sensitivity.
(b) 90\% C.L. upper limits on coupling constant $g^\prime$. Dark blue filled areas show the exclusion regions for $g^\prime$ at 90\% C.L., assuming the branching ratio $BR(Z^\prime\to \rm{invisible})$ predicted by the $L_\mu-L_\tau$ model; light blue areas are for $BR(Z^\prime\to \rm{invisible})=1$. The solid and dashed lines are the expected sensitivities for the two hypotheses. The red band shows the parameters region able to  explain the anomalous magnetic moment of the muon $(g-2)_\mu \pm 2\sigma$.}
\label{fig:Zprime}
\end{figure}

The interesting events are selected as those with exactly two tracks identified as muons coming from the interaction region. The recoil momentum is required to point within the ECL barrel  in order to exclude low-efficiency regions where photons from radiative processes can escape the detection and mimic the signal. This latter requirement is applied only for low recoil masses ($M_{rec}<3$ GeV/c$^2$), as contributions from radiative backgrounds are unlikely at higher masses.
Events with reconstructed photons above a minimum energy in the recoil direction are discarded. The transverse momentum of the dimuon system is required to be above some threshold, as this selection is very effective against $\mu^+\mu^-(\gamma)$ and $e^+e^-\mu^+\mu^-$ backgrounds. Finally, an additional selection to further suppress $\tau^+\tau^-$ events, exploits the fact that the $Z^\prime$ production is a final-state-radiation process from a final state muon,  while for the tau background the missing momentum comes from both $\tau$ decay neutrinos. For this purpose, a linear cut in the plane defined by the transverse momentum with respect to the higher and lower momentum leptons is applied \cite{Adachi:2019otg}.
After the final selection, almost all background sources with the exception of $e^+e^-\to e^+e^-\mu^+\mu^-$ are rejected.
Control samples on data are used to check the background yields expected from Monte Carlo samples and to estimate correction factors and related uncertainties. The largest sources of systematic uncertainty result from poor agreement between data and Monte Carlo simulations and limited statistics of the control samples, which are expected to reduce as the data size will increase.
A counting technique within mass windows with a size equal to twice the recoil mass resolution has been performed.
No anomalies have been observed, with all results below 3$\sigma$ local significance. 
The 90\% credibility level (C.L.) upper limits on the cross section have been computed using a Bayesian procedure, and results have been translated into 90\% C.L. upper limits on the  $g^\prime$ coupling constant. This latter is shown in Figure~\ref{fig:Zprime_up}.
Upper limits on $g^\prime$ stay in the range $[5 \times 10^{-2}-1]$  for $M_{Z^\prime} \leq 6$ GeV/c$^2$.

\begin{figure}
\centering
\includegraphics[width=0.50\textwidth]{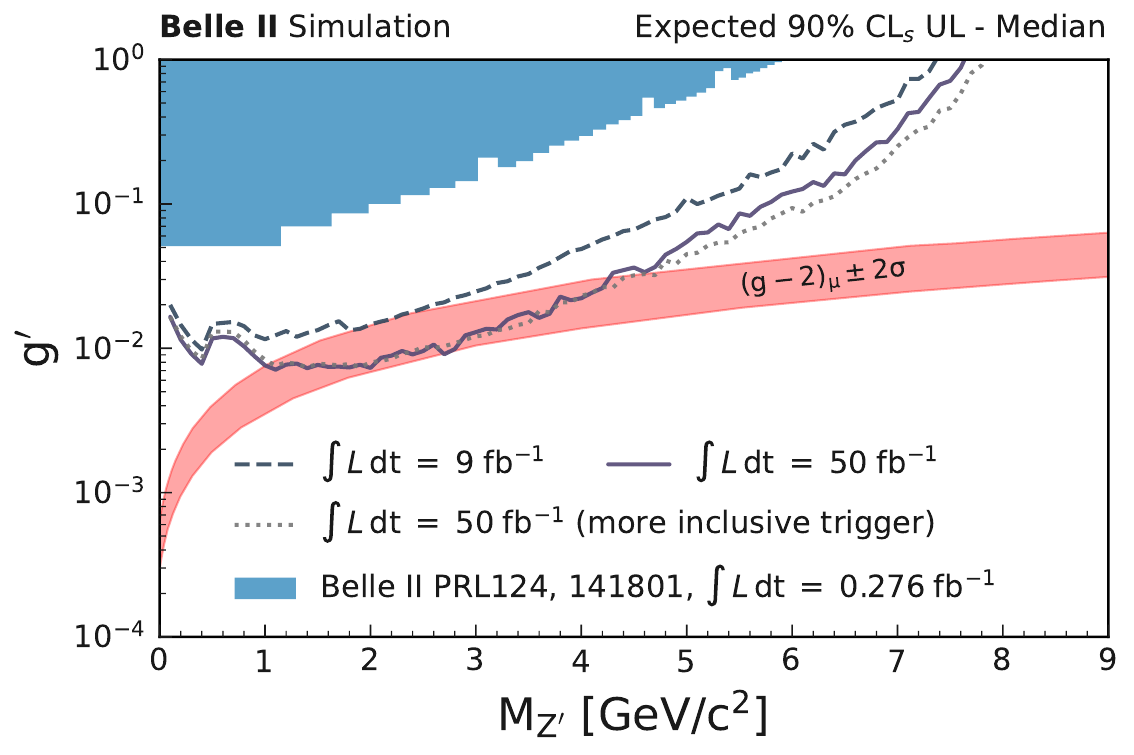}
\caption{ 90\% C.L. expected upper limits to the coupling constant $g^\prime$ for  integrated luminosities of $9\,\rm{fb^{-1}}$ and $50\,\rm{fb^{-1}}$. For the $50\,\rm{fb^{-1}}$ case, the effect of using the more inclusive trigger line is
compared. Note that a branching ratio $BR(Z^\prime\to \rm{invisible})=1$ was assumed. The current 90\% C.L. upper limits set with the 2018 data are shown too.}
\label{fig:Zprime_phase3}
\end{figure}

Updates of the analysis are expected to improve the sensitivity to $g^\prime$ by more than one order of magnitude, already with the integrated luminosity collected so far $\mathcal{O}(50\,\rm{fb}^{-1})$.
Figure~\ref{fig:Zprime_phase3} shows the expected 90\% C.L. upper limits on the $Z^\prime$ coupling constant $g^\prime$ for the case of 9 and 50 $\rm{fb}^{-1}$ with improvement in the analysis with respect to the 2018 data analysis, due to a better particle identification algorithm (which makes use of the KLM detector), a better vertex resolution and the use of Multivariate Analysis (MVA) selection criteria. It is further shown the effect of using a more inclusive trigger available from the 2020 data taking period, which allows improving the sensitivity  especially in the high mass region. Figure~\ref{fig:Zprime_phase3} shows the possibility to exclude a good part of the region of the parameters able to explain the $(g-2)_\mu \pm 2\sigma$ anomaly already with an integrated luminosity of $\mathcal{O}(50\,\rm{fb^{-1}})$.

\section{Lepton Flavour Violating $Z^\prime$}

As an extension of the above search for the invisibly decaying $Z^\prime$, the existence of a Lepton Flavour Violating (LFV) $Z^\prime$ boson  has been investigated too. Specifically, this search focused on an invisibly decaying LFV $Z^\prime$ produced in the process $e^+e^-\to e^\pm \mu^\mp Z^\prime,\,Z^\prime\to \rm{invisible}$ \cite{Adachi:2019otg}.
The experimental signature consists of two oppositely charged tracks with different flavours plus missing energy. 
Due to the non-availability of a robust signal generator, a model-independent search has been performed by using the same selection criteria of the above ($L_\mu-L_\tau$ model) $Z^\prime$ search, aside from the obvious requirement to have an electron instead of a muon in the final state. The search for a LFV final state is expected to have a small SM background: the only background contribution after the final selection is $ e^+e^-\to\tau^+\tau^-(\gamma)$ with taus going to one-prong muon and one-prong electron. In the case of signal events, a bump in the distribution of the mass recoiling against the $e\mu$ system is expected at the mass value of the LFV $Z^\prime$. 

This search has been performed with the 2018 data-set, which is the same as that used for the $Z^\prime$ search in the $L_\mu-L_\tau$ framework.
Also in that case, no anomalies have been observed above 3$\sigma$ local significance.

\begin{figure}
\centering
\includegraphics[width=0.48\textwidth]{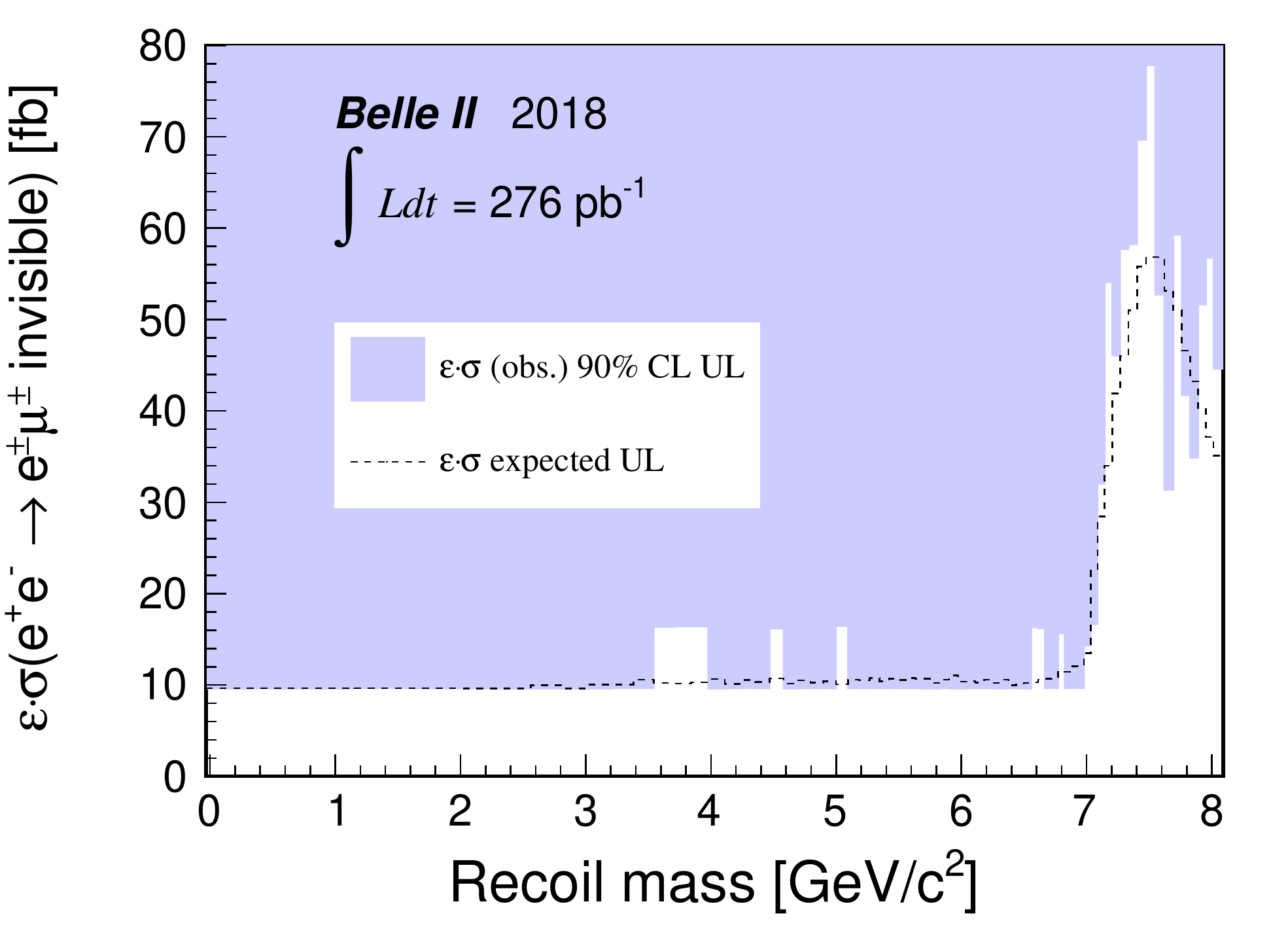}
\caption{ Results for the  $e^+e^-\to e^\pm\mu^\mp Z^\prime,\,Z^\prime\to \rm{invisible}$   search performed with the 2018 dataset (276 pb$^{-1}$) \cite{Adachi:2019otg}: 90\% C.L. upper limits on  signal efficiency times cross section $\epsilon\times\sigma(e^+e^-\to e^\pm\mu^\mp\,\rm{invisible})$ (light blue
areas). The dashed line represents the expected sensitivity.}
\label{fig:LFV_Zprime_up}
\end{figure}

As a robust signal generator was not available,  model-independent 90\% C.L. upper limits on the LFV $Z^\prime$ efficiency times cross section have been computed by using a Bayesian procedure, as shown in Figure~\ref{fig:LFV_Zprime_up}.

\section{Dark Photon}

One of the most attractive dark sector models introduces a new spontaneously broken Abelian gauge group $U(1)^\prime$, with a new massive gauge boson called $\textit{dark photon}$ $A^\prime$, which mirrors the SM hypercharge interaction. 
In the minimal scenario, it does not couple directly to charged SM particles, but the coupling is done via a kinetic mixing mechanism to the SM photon with  strength $\epsilon<1$ \cite{Holdom:1985ag,Pospelov:2007mp,FAYET1990743}. The mixing strength can be understood as the suppression factor relative to the coupling with the elementary charge $e$.

\subsection{Radiative Processes}
\begin{figure}
  \centering
    \includegraphics[width=0.48\textwidth]{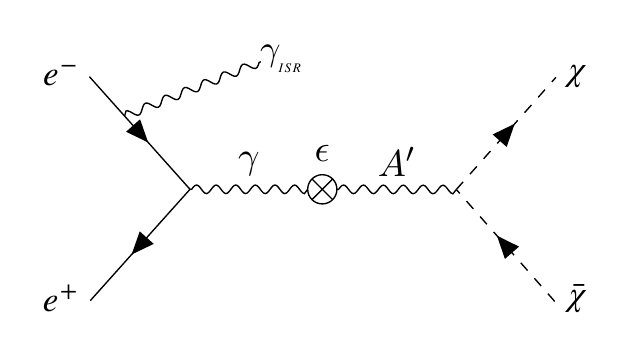}
    \caption{Feynman diagram for a dark photon produced by kinematic mixing in a radiative process and decaying invisibly.}
  \label{fig:Aprime_feynman}
\end{figure}

A simple way to search for a dark photon at an $e^+e^-$ collider is in the initial-state-radiation (ISR) process $e^+e^-\to\gamma_{\rm{ISR}}\,A^\prime$, which is shown in Figure~\ref{fig:Aprime_feynman}. The related cross section is proportional to $\epsilon^2\alpha^2/s$ where $\alpha$ is the electromagnetic coupling. 

The dark photon decay modes depend on its mass and coupling, and on the presence of any other candidate of the dark sector. 
It can decay into SM final states via $A^\prime\to l^+l^-$ or $A^\prime\to h^+h^-$ ($l$=leptons, $h$=hadrons) with an expected branching ratio equal to that of a virtual photon of mass $m_{A^\prime}$ \cite{Batell:2009yf}.
In such a scenario, the dark photon search proceeds by looking for a resonance in the invariant mass distribution of the reconstructed daughter particles. In the case of long-lived $A^\prime$, the decay would occur far from the interaction point; thus one expects that the two-tracks vertex is  displaced with respect to the interaction region.

If a sufficiently low mass dark matter state $\chi$ exists (such that $m\chi<1/2\,m_{A^\prime}$), one can assume that the dominant decay mode of the dark photon is into DM via $A^\prime \to \chi\bar{\chi}$. As the probability of DM to interact with the detector is negligible, this latter case is usually referred to as \textit{invisible} decay, while the former case  as \textit{visible} decay. The experimental signature of the process $e^+e^-\to\gamma_{\rm{ISR}}\,A^\prime,\, A^\prime\to \chi\bar{\chi}$  would be only a mono–energetic ISR photon, along with a significant missing energy. In such a case, the energy of the ISR photon is related to the dark photon mass $m_{A^\prime}$ through the relationship  $E_\gamma=\frac{s-m^2_{A^\prime}}{2\sqrt{s}}$.
Thus, the search occurs as a scan of the  squared mass distribution of the recoiling system against the ISR photon.

Two recent searches by the BaBar experiments looked for the $A^\prime$  production in both visible \cite{PhysRevLett.113.201801} and invisible final states \cite{PhysRevLett.119.131804}.

\subsubsection{Invisible Decay:}\label{sec:SP}
Searches for a dark photon in the ISR process are expected also at Belle II \cite{10.1093/ptep/ptz106}. In particular, in the invisible decay channel, setting new stringent upper limits is already possible with the data collected so far.

A simulation of the expected backgrounds and signal efficiencies with the Belle~II detector was carried out, and the sensitivity to dark photons decaying into light DM was determined. In the computation, the $A^\prime$ was assumed to decay predominantly to dark matter, resulting in a 100\% branching fraction to invisible.
Signal events have been generated with the \textsc{MadGraph5} generator, for several dark photon mass hypotheses.
The backgrounds are mainly due to the high cross section QED processes, namely $e^+e^-\to e^+e^-\gamma(\gamma)$ (generated with \textsc{BHWIDE} and \textsc{TEEGG} \cite{Jadach:1995nk,Karlen:1987vk}) and $e^+e^-\to \gamma\gamma(\gamma)$ (generated with \textsc{Babayaga@nlo}). Their contributions come out when all but one photon are undetected, being out of the acceptance, or due to gaps or inefficient regions. Therefore a very good knowledge of the detector efficiencies is needed for such analysis.

\begin{figure}
  \centering
    \includegraphics[width=0.5\textwidth]{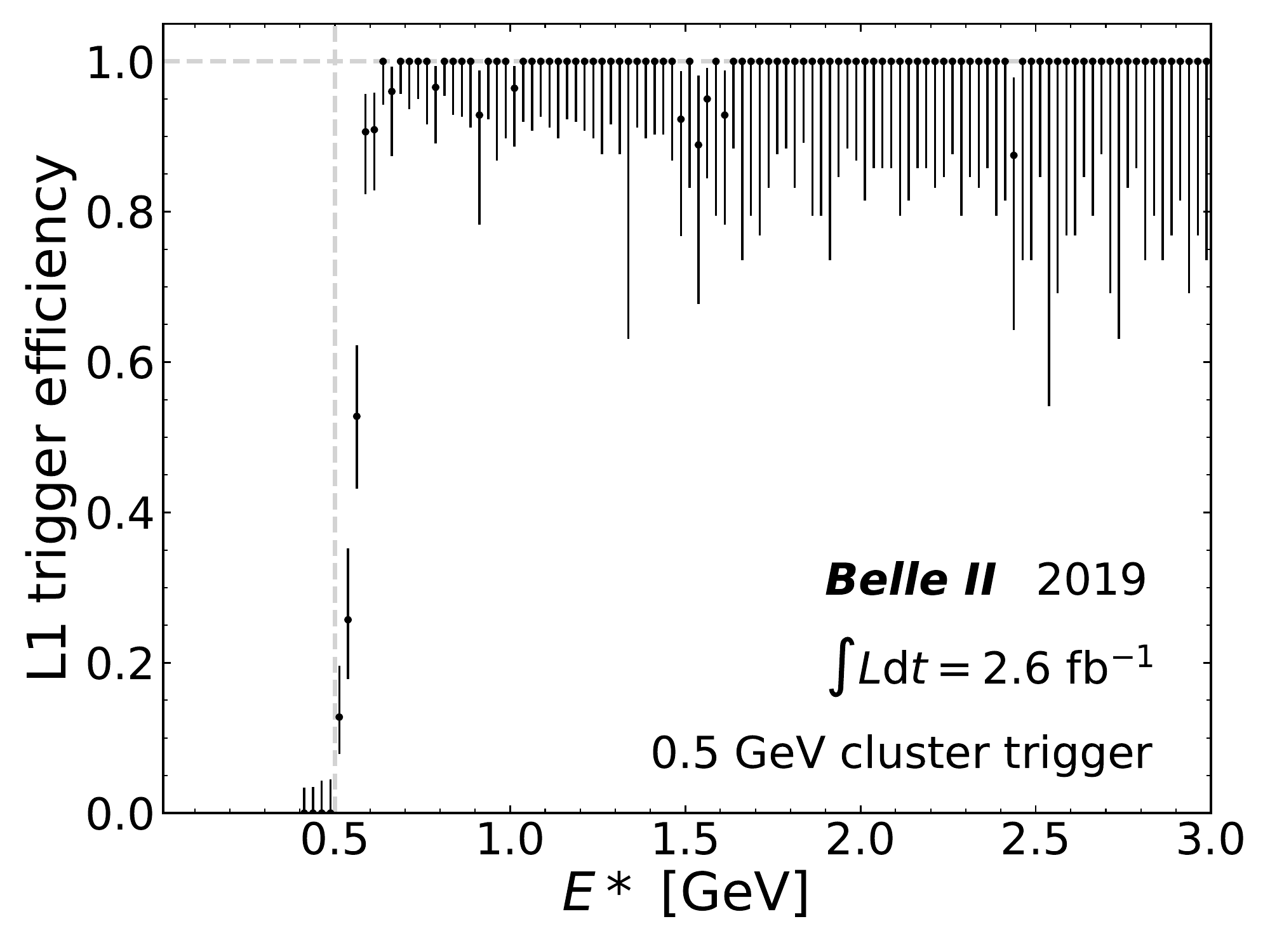}
\caption{ Efficiency of the Level--1  single-photon trigger with a 0.5 GeV threshold as a function of energy measured by using a radiative muon pair sample.}
\label{fig:trigger}
\end{figure}

This analysis requires the implementation of a dedicated first level (L1) trigger sensitive to single photons. It was not available at the predecessor experiment Belle, and only partially available at the BaBar experiment (BaBar recorded only 53 fb$^{-1}$ of data with the single-photon trigger on). 
The main difficulty in realizing such a trigger is due to the high rate, mainly due to radiative Bhabba and $e^+e^-\to\gamma\gamma$ events where only a single photon is produced within the detector acceptance.
Belle II implemented several single-photon trigger lines already from the first operations, with a trigger rate well below the L1 system design maximum output rate of 30 kHz \cite{Iwasaki:2011za}.
Remarkably good efficiency values have been measured on the data collected so far, as shown in Figure~\ref{fig:trigger}. 

The analysis selection requires a single energetic ECL cluster  and no other cluster above a minimum energy threshold as well as tracks above a minimum track momentum. 
In order to suppress backgrounds due to ECL photon inefficiency, events with KLM clusters not matched with the ECL are discarded.
Furthermore, a final selection based on a multivariate cut in the ISR photon energy (in the CMS) versus the polar angle (in the laboratory frame) plane is performed \cite{10.1093/ptep/ptz106}.

The expected sensitivity to the mixing parameter $\epsilon$ with an integrated luminosity of 20 fb$^{-1}$ is shown in  Figure~\ref{fig:invisi_dp}. Belle II is already able to set a limit considerably lower than the equivalent of BaBar \cite{PhysRevLett.119.131804}.
The Belle~II better performances with respect to BaBar is mainly due to a more homogeneous electromagnetic calorimeter, with no projective gaps to the interaction point.  In BaBar, a large irreducible background was due to  $e^+e^-\to \gamma\gamma(\gamma)$  events, with due to one or more photons escaping the detection by passing through projective cracks between adjacent crystals. 
An additional improvement comes from SuperKEKB beam lower energies, which provide a smaller boost and larger electromagnetic calorimeter angular coverage, allowing for a larger acceptance of signal events. 
Furthermore, the  Belle II  $K_L$ and muon  detector allows to veto photons  not detected by the electromagnetic calorimeter. 

The sensitivity to the high mass range is limited by the energy threshold for the single-photon trigger, which, for such a study, has been here conservatively assumed to be 1.8~GeV. However, based on the use of trigger with lower energy thresholds, a significant improvement in sensitivity in the high mass region is expected.

\subsubsection{Visible Decay:}

\begin{figure}
\begin{subfigure}{.5\textwidth}
  \centering
    \includegraphics[width=0.950\textwidth]{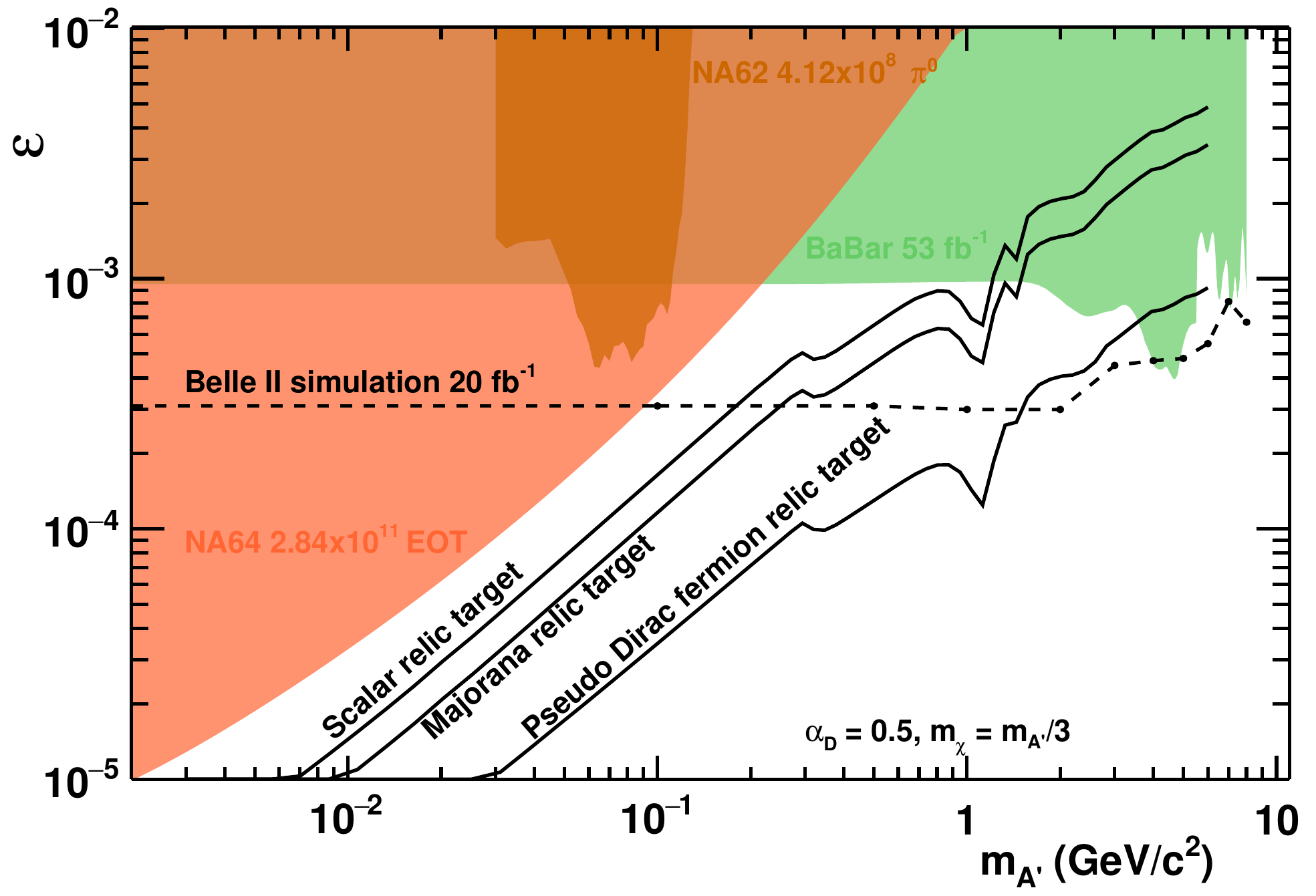}
    \caption{}
  \label{fig:invisi_dp}
\end{subfigure}%
\begin{subfigure}{.5\textwidth}
  \centering
    \includegraphics[width=0.950\textwidth]{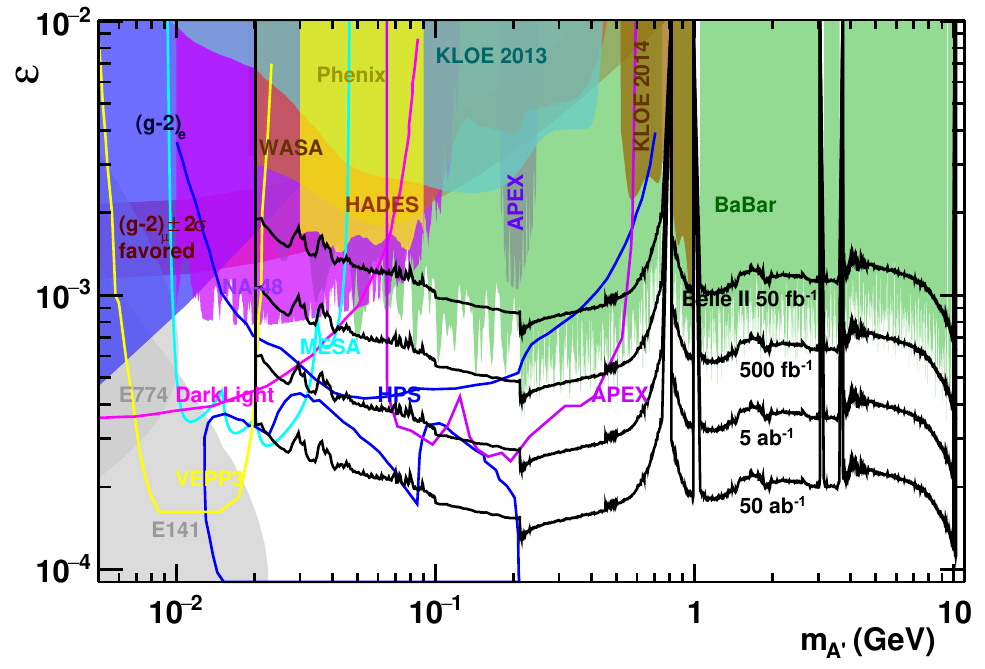}
    \caption{}
  \label{fig:vis_dp}
\end{subfigure}
\caption{ (a) Expected sensitivity to the kinetic mixing strength in the invisible dark photon search through the process  $e^+e^-\to\gamma_{\rm{ISR}}\,A^\prime,\, A^\prime\to \chi\bar{\chi}$. Current bounds are from BaBar \cite{PhysRevLett.119.131804} and NA64 \cite{NA64:2019imj}. Solid black lines show the favored dark photon parameter regions, which account for the observed relic DM density for the scalar, pseudo-Dirac, and Majorana DM. (b) Projected limits to the dark photon kinetic mixing parameter $\epsilon$  via the visible decay process  $e^+e^-\to\gamma_{\rm{ISR}}\,A^\prime,\, A^\prime\to l\bar{l}$ with the Belle~II experiment, compared to other experiments \cite{10.1093/ptep/ptz106}. Strongest constraints for $m_{A^\prime}\lesssim 10\, \rm{GeV}$ are from BaBar \cite{PhysRevLett.113.201801}, electron  anomalous  magnetic momentum measurements \cite{Endo:2012hp}, KLOE \cite{Babusci:2014sta}, NA48 \cite{Batley:2015lha} and other beam-dump experiments \cite{Andreas:2012mt,Blumlein:2011mv}.}
\end{figure}

Searches for the dark photon visible decays at Belle~II are also planned. These are rather more experimentally challenging  with respect to the invisible case as they suffer for  high SM backgrounds.
Preliminary studies for the visible search at Belle II have been reported in Ref.~\cite{10.1093/ptep/ptz106}, where the  BaBar results  for the $A^\prime$ visible decay  via  $e^+e^-$ and $\mu^+\mu^-$ final states \cite{PhysRevLett.113.201801} have been used to compute expected sensitivities in Belle~II. BaBar searched for a narrow peak in the di-lepton mass spectrum over the background, mainly from QED processes. Taking into account several factors of improvement of Belle~II with respect to BaBar, namely the better mass resolution ($\sim$ a factor 2) due to the larger drift chamber radius and the better vertex detector, and better trigger efficiency for both muons ($\sim$ a factor 1.1) and electrons ($\sim$ a factor 2), the projected sensitivities for several values of integrated luminosity are shown in Figure~\ref{fig:vis_dp}. 

Belle II results are expected to be competitive with BaBar once an integrated luminosity of  $\mathcal{O}(500\,\rm{fb^{-1}})$ has been accumulated, while with the target luminosity of 50~ab$^{-1}$, the expectation is to constrain $\epsilon$ down to $\mathcal{O}(10^{-4})$.

\subsection{Dark Higgsstrahlung process}

\begin{figure}
  \centering
    \includegraphics[width=0.48\textwidth]{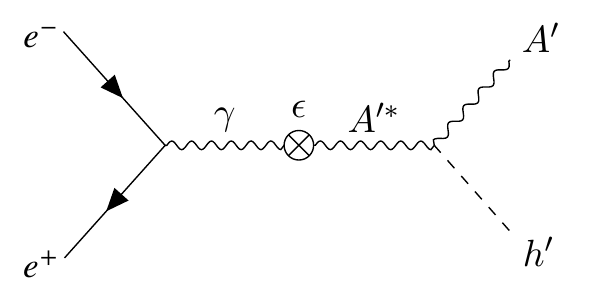}
    \caption{Feynman diagram for the dark Higgsstrahlung process.}
  \label{fig:DH_feynman}
\end{figure}

The dark photon mass could be generated, in close analogy with the SM, via a spontaneous breaking mechanism of the dark $U(1)^\prime$ gauge group, by introducing a Higgs-like particle $h^\prime$, called \textit{dark Higgs} \cite{Batell:2009yf}.
The dark Higgsstrahlung process $e^+e^- \to A^*\to A^\prime\, h^\prime$, with $A^\prime$ decaying into lepton or hadron pairs, is then an interesting reaction to search for both the $A^\prime$ and the $h^\prime$ at an  $e^+e^-$ collider, as shown in Figure~\ref{fig:DH_feynman}. Its cross section is proportional to the product $ \epsilon^2 \times \alpha_{\rm{D}}$, where $\alpha_{\rm{D}}$ is the unknown dark sector coupling constant,  and depends on the boson masses \cite{Batell:2009yf}.

Two different scenarios can happen depending on the dark photon  and dark Higgs masses ($m_{A^\prime}$, $m_{h^\prime}$).  For $m_{h^\prime}>2m_{A^\prime}$, the dark Higgs  would promptly decay to a dark photon pair, thus resulting in a six charged particles final state. This scenario was investigated by Belle \cite{TheBelle:2015mwa} and BaBar \cite{Lees:2012ra}.
The case with $m_{A^\prime}<m_{h^\prime}<2m_{A^\prime}$  is similar to the previous one, with one dark photon is off-shell. 
On the contrary, if $h^\prime$ is lighter than the dark photon, it would be long-lived for most of the parameter phase space, thus escaping the detection (\textit{invisible dark Higgs} scenario). This latter case was investigated by the KLOE experiment  only, for $A^\prime$ masses up to $\simeq$ 1 GeV \cite{Babusci:2015zda}.

\subsubsection{Invisible Dark Higgs:}
Being less constrained, Belle II is planning to search for this latter scenario, i.e., the $A^\prime$ and $h^\prime$ production  via the $e^+e^- \to A^\prime h^\prime,\, A^\prime\to\mu^+\mu^-,\,h^\prime\to \rm{invisible}$ process by using the data collected during 2019, which correspond to an integrated luminosity of $\sim9\,\rm{fb}^{-1}$. 

The final state is given by a pair of opposite charge muons plus missing energy. In the case of signal events, the presence of simultaneous peaks both in the distribution of the dimuon invariant mass $M_{\mu\mu}$ and in the distribution of the invariant mass $M_{\rm{rec}}$ of the system recoiling against the two muons is expected. 
The interesting phase space region has a triangular shape, being limited on the left by the required condition $M_{\rm{rec}} <  M_{\mu\mu}$ (being $m_{h^\prime}<$ $m{_A^\prime}$) and on the right by energy conservation:  $M_{\rm{rec}} +  M_{\mu\mu} < \sqrt{s}$. The measurement is thus performed in the range $2 m_\mu <  m_{A^\prime}  < 10.58$ GeV/c$^2$ with the aforementioned constraints, corresponding to a sizeable enlargement of the parameters region explored by KLOE. 

At the moment, a full Monte Carlo study of the expected backgrounds and signal efficiencies with the Belle II detector was carried out. Signal samples have been generated with the \textsc{MadGraph5} generator for different $A^\prime$ and $h^\prime$ mass hypotheses, covering all the interesting mass range with a step corresponding to the mass resolution in the two directions.
Concerning the background, main sources are the same as for the \textit{invisible} $Z^\prime$ analysis, namely $e^+e^- \rightarrow \mu^+\mu^-(\gamma)$,  $e^+e^- \rightarrow \tau^+\tau^-(\gamma)$ and $e^+e^- \rightarrow e^+e^-\mu^+\mu^-$, and many of the analysis selections are then very similar.

The interesting events are selected as those with exactly two muon tracks coming from the interaction region. The recoil momentum is required to point within the ECL barrel region in order to exclude regions where photons from radiative processes can escape the detection. Furthermore, events with a reconstructed photon in the recoil direction are discarded. The transverse momentum of the dimuon system is required to be above some threshold, as this selection is very effective against $\mu^+\mu^-(\gamma)$ and $e^+e^-\mu^+\mu^-$ backgrounds. A further suppression has been implemented based on the helicity angle, defined as the angle between one of the two muons (e.g. $\mu^-$) in the dimuon rest frame and the dimuon flight direction in the CM system, which takes into account the different dimuon production processes in case of signal or background events. Interesting events in data are selected by using the two-track CDC trigger, which requires at least two tracks with an opening angle between the two candidates in the transverse plane exceeding 90$^\circ$.

\begin{figure}
  \centering
    \includegraphics[width=0.58\textwidth]{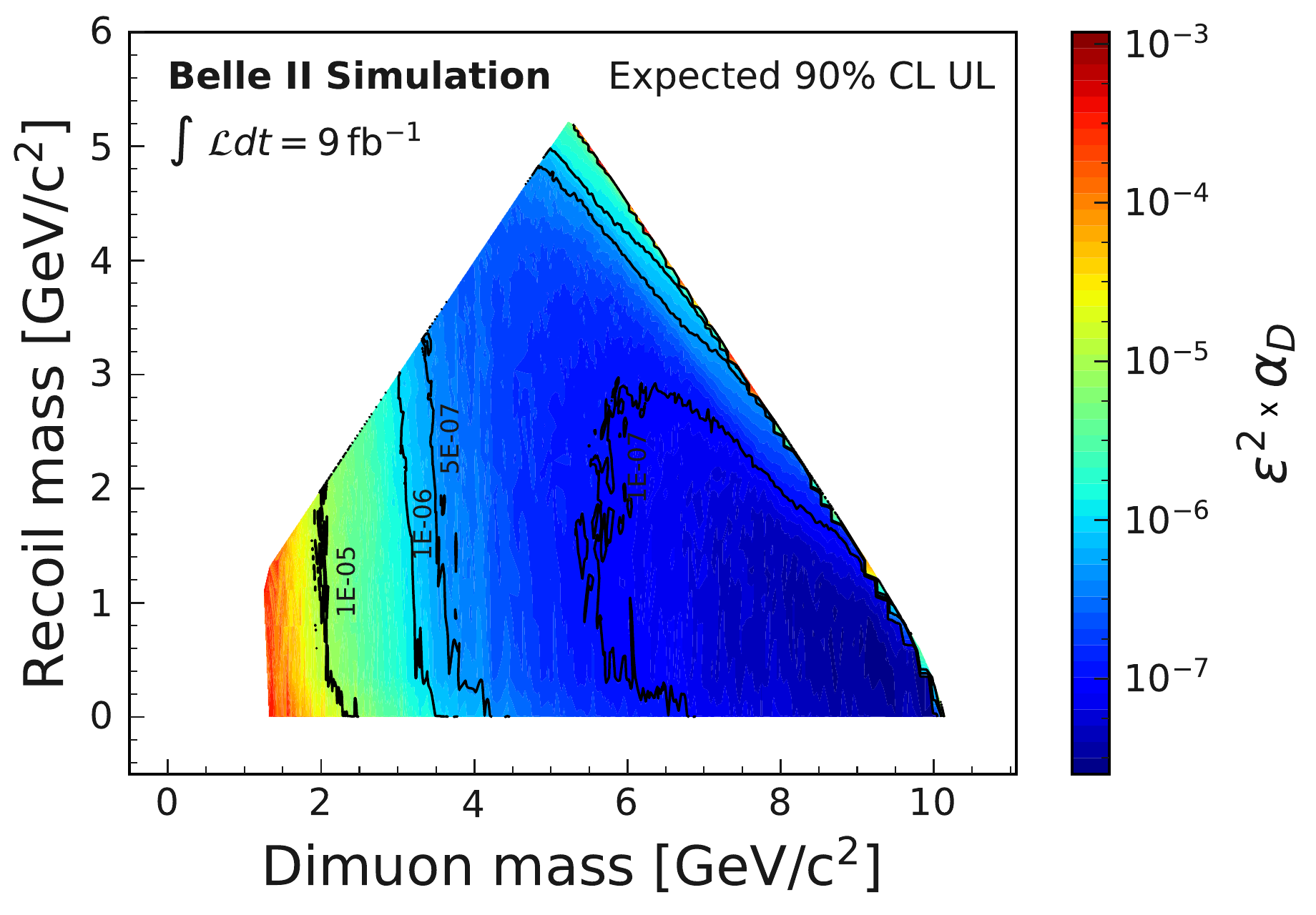}
    \caption{90\% C.L. expected upper limits to the coupling constant product $\epsilon^2\times\alpha_D$ via the process $e^+e^- \to A^\prime h^\prime,\, A^\prime\to\mu^+\mu^-,\,h^\prime\to \rm{invisible}$, with an integrated luminosity of $9\,\rm{fb}^{-1}$, which corresponds to about the 2019 data-taking integrated luminosity.}
  \label{fig:DH}
\end{figure}

The analysis strategy relies on a mass scan plus counting technique within mass windows of size proportional to the experimental resolution into the two-dimensional mass phase space. 
Figure~\ref{fig:DH} shows the expected 90\% C.L. upper limits to the coupling constant product $\epsilon^2\times\alpha_D$. 
With the data-set collected during  2019, Belle~II is expected to constrain $\epsilon^2\times\alpha_D$ down to $\mathcal{O}(10^{-7})$ in most of the phase space. 
The sensitivity in the low mass region ($m_{A^\prime}\lesssim 4\,\rm{GeV/c^2}$) is strongly limited by the poor trigger efficiency for small opening angle muons. In that region, a noticeable improvement is expected  with the use of more inclusive triggers available from 2020 on (e.g. the KLM based single-muon trigger): it will allow to fully recover this region and to achieve a sensitivity comparable to that in the high mass region.

\section{Conclusions}

This paper provides an overview of the Belle~II experiment and its capabilities in performing low-mass dark sector searches. 

The first results obtained with the data collected during the 2018 data-taking period have been reviewed. Specifically, the searches for an ALP decaying into two photons and for an invisibly decaying $Z^\prime$ (in the framework of the $L_\mu-L_\tau$ model, as well as the model-independent case of a LFV $Z^\prime$)   have been presented.  
These results are already state of the art and can be furthermore updated in the future due to the expected increased data-set and the usage of more inclusive trigger lines, which have been recently activated.

In addition to the aforementioned searches, the expected sensitivity  with the first or short-term collected data-sets has been shown for different searches: ALP decaying invisibly,   dark photon searches in radiative processes (both the visible and invisible decay channel) and the dark Higgsstrahlung. 

Beyond the searches mentioned here, the dark sector physics program at Belle~II is wild and still includes many other studies, such as dark scalars, long-lived particles and magnetic monopoles.

\section*{References}
\bibliography{bibliography}

\providecommand{\newblock}{}
\begin{thebibliography}{10}
\expandafter\ifx\csname url\endcsname\relax
  \def\url#1{{\tt #1}}\fi
\expandafter\ifx\csname urlprefix\endcsname\relax\def\urlprefix{URL }\fi
\providecommand{\eprint}[2][]{\url{#2}}

\bibitem{refId0}
Ade P~A~R {\em et~al.\/} (Planck) 2016 {\em Astron. Astrophys.\/} {\bf 594} A13

\bibitem{1483008}
Essig R, Jaros J~A and Wester W 2013 {Dark Sectors and New, Light,
  Weakly-Coupled Particles} (\textit{Preprint} \eprint{1311.0029})

\bibitem{Batell:2009di}
Batell B, Pospelov M and Ritz A 2009 {\em Phys. Rev. D\/} {\bf 80} 095024
  (\textit{Preprint} \eprint{0906.5614})

\bibitem{Adriani:2008zr}
Adriani O {\em et~al.\/} (PAMELA) 2009 {\em Nature\/} {\bf 458} 607--609

\bibitem{FermiLAT:2011ab}
Ackermann M {\em et~al.\/} (Fermi-LAT) 2012 {\em Phys. Rev. Lett.\/} {\bf 108}
  011103

\bibitem{PhysRevLett.110.141102}
Aguilar M {\em et~al.\/} (AMS) 2013 {\em Phys. Rev. Lett.\/} {\bf 110}(14)
  141102

\bibitem{Bennett:2006fi}
Bennett G~W {\em et~al.\/} (Muon g-2) 2006 {\em Phys. Rev. D\/} {\bf 73} 072003

\bibitem{Boveia:2018yeb}
Boveia A and Doglioni C 2018 {\em Ann. Rev. Nucl. Part. Sci.\/} {\bf 68}
  429--459

\bibitem{Borodatchenkova:2005ct}
Borodatchenkova N, Choudhury D and Drees M 2006 {\em Phys. Rev. Lett.\/} {\bf
  96} 141802 (\textit{Preprint} \eprint{hep-ph/0510147})

\bibitem{Fayet:2007ua}
Fayet P 2007 {\em Phys. Rev. D\/} {\bf 75} 115017 (\textit{Preprint}
  \eprint{hep-ph/0702176})

\bibitem{Batell:2009yf}
Batell B, Pospelov M and Ritz A 2009 {\em Phys. Rev. D\/} {\bf 79} 115008
  (\textit{Preprint} \eprint{0903.0363})

\bibitem{Essig:2009nc}
Essig R, Schuster P and Toro N 2009 {\em Phys. Rev. D\/} {\bf 80} 015003
  (\textit{Preprint} \eprint{0903.3941})

\bibitem{Essig:2013vha}
Essig R, Mardon J, Papucci M, Volansky T and Zhong Y~M 2013 {\em JHEP\/} {\bf
  11} 167

\bibitem{Reece:2009un}
Reece M and Wang L~T 2009 {\em JHEP\/} {\bf 07} 051 (\textit{Preprint}
  \eprint{0904.1743})

\bibitem{Yin:2016hbi}
Yin P~F and Zhu S~H 2016 {\em Front. Phys. (Beijing)\/} {\bf 11} 111403

\bibitem{Alexander:2016aln}
Alexander J {\em et~al.\/} 2016 {Dark Sectors 2016 Workshop: Community Report}
  (\textit{Preprint} \eprint{1608.08632})

\bibitem{Agrawal:2021dbo}
Agrawal P {\em et~al.\/} 2021  (\textit{Preprint} \eprint{2102.12143})

\bibitem{Battaglieri:2017aum}
Battaglieri M {\em et~al.\/} 2017 {US Cosmic Visions: New Ideas in Dark Matter
  2017: Community Report} {\em {U.S. Cosmic Visions: New Ideas in Dark
  Matter}\/} (\textit{Preprint} \eprint{1707.04591})

\bibitem{AKAI2018188}
Akai K, Furukawa K and Koiso H 2018 {\em Nucl. Instrum. Meth. A\/} {\bf 907}
  188 -- 199

\bibitem{Bona:2007qt}
Bona M {\em et~al.\/} (SuperB) 2007

\bibitem{ABASHIAN2002117}
Abashian A {\em et~al.\/} (Belle) 2002 {\em Nucl. Instrum. Meth. A\/} {\bf 479}
  117 -- 232

\bibitem{10.1093/ptep/ptz106}
Kou E {\em et~al.\/} (Belle-II) 2019 {\em PTEP\/} {\bf 2019}

\bibitem{Abudin_n_2020}
Abudin{\'{e}}n F,  {\em et~al.\/} (Belle-II) 2020 {\em Chinese Physics C\/}
  {\bf 44} 021001

\bibitem{BelleII:2020fag}
Abudin\'en F {\em et~al.\/} (Belle-II) 2020 {\em Phys. Rev. Lett.\/} {\bf 125}
  161806

\bibitem{Adachi:2019otg}
Adachi I {\em et~al.\/} (Belle-II) 2020 {\em Phys. Rev. Lett.\/} {\bf 124}
  141801 (\textit{Preprint} \eprint{1912.11276})

\bibitem{Agostinelli:2002hh}
Agostinelli S {\em et~al.\/} (GEANT4) 2003 {\em Nucl. Instrum. Meth. A\/} {\bf
  506} 250--303

\bibitem{Kuhr:2018lps}
Kuhr T, Pulvermacher C, Ritter M, Hauth T and Braun N (Belle-II Framework
  Software Group) 2019 {\em Comput. Softw. Big Sci.\/} {\bf 3} 1
  (\textit{Preprint} \eprint{1809.04299})

\bibitem{jaeckel}
Jaeckel J and Ringwald A 2010 {\em Ann. Rev. Nucl. Part. Sci.\/} {\bf 60}
  405--437

\bibitem{Peccei}
Peccei R~D and Quinn H~R 1977 {\em Phys. Rev. Lett.\/} {\bf 38}(25) 1440--1443

\bibitem{Nomura:2008ru}
Nomura Y and Thaler J 2009 {\em Phys. Rev. D\/} {\bf 79} 075008
  (\textit{Preprint} \eprint{0810.5397})

\bibitem{Arias:2012az}
Arias P, Cadamuro D, Goodsell M, Jaeckel J, Redondo J and Ringwald A 2012 {\em
  JCAP\/} {\bf 06} 013 (\textit{Preprint} \eprint{1201.5902})

\bibitem{Cicoli:2014bfa}
Cicoli M, Conlon J~P, Marsh M~C~D and Rummel M 2014 {\em Phys. Rev. D\/} {\bf
  90} 023540 (\textit{Preprint} \eprint{1403.2370})

\bibitem{Jaeckel:2014qea}
Jaeckel J, Redondo J and Ringwald A 2014 {\em Phys. Rev. D\/} {\bf 89} 103511
  (\textit{Preprint} \eprint{1402.7335})

\bibitem{torben}
Dolan M, Ferber T, Hearty C, Kahlhoefer F and Schmidt-Hoberg K 2017 {\em
  JHEP\/} {\bf 2017}

\bibitem{Balossini:2008xr}
Balossini G, Bignamini C, Calame C~M~C, Montagna G, Nicrosini O and Piccinini F
  2008 {\em Phys. Lett. B\/} {\bf 663} 209--213 (\textit{Preprint}
  \eprint{0801.3360})

\bibitem{Czyz:2017veo}
Czy\.z H, Kisza P and Tracz S 2018 {\em Phys. Rev. D\/} {\bf 97} 016006
  (\textit{Preprint} \eprint{1711.00820})

\bibitem{Alwall:2014hca}
Alwall J, Frederix R, Frixione S, Hirschi V, Maltoni F, Mattelaer O, Shao H~S,
  Stelzer T, Torrielli P and Zaro M 2014 {\em JHEP\/} {\bf 07} 079
  (\textit{Preprint} \eprint{1405.0301})

\bibitem{Li:2018qnh}
Li Q and Yan Q~S 2018  (\textit{Preprint} \eprint{1804.00125})

\bibitem{ZERNIKE1934689}
von F~Zernike 1934 {\em Physica\/} {\bf 1} 689--704 ISSN 0031-8914
  \urlprefix\url{https://www.sciencedirect.com/science/article/pii/S0031891434802595}

\bibitem{Knapen:2016moh}
Knapen S, Lin T, Lou H~K and Melia T 2017 {\em Phys. Rev. Lett.\/} {\bf 118}
  171801 (\textit{Preprint} \eprint{1607.06083})

\bibitem{Banerjee:2020fue}
Banerjee D {\em et~al.\/} (NA64) 2020 {\em Phys. Rev. Lett.\/} {\bf 125} 081801
  (\textit{Preprint} \eprint{2005.02710})

\bibitem{Dobrich:2015jyk}
D\"obrich B, Jaeckel J, Kahlhoefer F, Ringwald A and Schmidt-Hoberg K 2016 {\em
  JHEP\/} {\bf 02} 018 (\textit{Preprint} \eprint{1512.03069})

\bibitem{Chudasama:2019jmm}
Chudasama R (CMS) 2019 {\em PoS\/} {\bf DIS2019} 084

\bibitem{Aloni:2019ruo}
Aloni D, Fanelli C, Soreq Y and Williams M 2019 {\em Phys. Rev. Lett.\/} {\bf
  123} 071801 (\textit{Preprint} \eprint{1903.03586})

\bibitem{He:1991qd}
He X~G, Joshi G~C, Lew H and Volkas R~R 1991 {\em Phys. Rev. D\/} {\bf 44}
  2118--2132

\bibitem{Shuve:2014doa}
Shuve B and Yavin I 2014 {\em Phys. Rev. D\/} {\bf 89} 113004

\bibitem{Altmannshofer:2016jzy}
Altmannshofer W, Gori S, Profumo S and Queiroz F~S 2016 {\em JHEP\/} {\bf 12}
  106

\bibitem{Baek:2001kca}
Baek S, Deshpande N~G, He X~G and Ko P 2001 {\em Phys. Rev. D\/} {\bf 64}
  055006 (\textit{Preprint} \eprint{hep-ph/0104141})

\bibitem{Ma:2001md}
Ma E, Roy D~P and Roy S 2002 {\em Phys. Lett. B\/} {\bf 525} 101--106
  (\textit{Preprint} \eprint{hep-ph/0110146})

\bibitem{Altmannshofer:2016brv}
Altmannshofer W, Chen C~Y, Bhupal~Dev P~S and Soni A 2016 {\em Phys. Lett. B\/}
  {\bf 762} 389--398 (\textit{Preprint} \eprint{1607.06832})

\bibitem{Crivellin:2015mga}
Crivellin A, D'Ambrosio G and Heeck J 2015 {\em Phys. Rev. Lett.\/} {\bf 114}
  151801 (\textit{Preprint} \eprint{1501.00993})

\bibitem{Baek:2017sew}
Baek S 2018 {\em Phys. Lett. B\/} {\bf 781} 376--382 (\textit{Preprint}
  \eprint{1707.04573})

\bibitem{Aaij:2013qta}
Aaij R {\em et~al.\/} (LHCb) 2013 {\em Phys. Rev. Lett.\/} {\bf 111} 191801
  (\textit{Preprint} \eprint{1308.1707})

\bibitem{Zhang:2020fiu}
Zhang Y, Yu Z, Yang Q, Song M, Li G and Ding R 2021 {\em Phys. Rev. D\/} {\bf
  103} 015008 (\textit{Preprint} \eprint{2012.10893})

\bibitem{PhysRevD.94.011102}
Lees J~P,  {\em et~al.\/} (BaBar) 2016 {\em Phys. Rev. D\/} {\bf 94}(1) 011102

\bibitem{Sirunyan:2018nnz}
Sirunyan A~M {\em et~al.\/} (CMS) 2019 {\em Phys. Lett. B\/} {\bf 792} 345--368

\bibitem{Curtin:2014cca}
Curtin D, Essig R, Gori S and Shelton J 2015 {\em JHEP\/} {\bf 02} 157

\bibitem{Jadach:1999vf}
Jadach S, Ward B~F~L and Was Z 2000 {\em Comput. Phys. Commun.\/} {\bf 130}
  260--325 (\textit{Preprint} \eprint{hep-ph/9912214})

\bibitem{Berends:1984gf}
Berends F~A, Daverveldt P~H and Kleiss R 1985 {\em Nucl. Phys. B\/} {\bf 253}
  441--463

\bibitem{Holdom:1985ag}
Holdom B 1986 {\em Phys. Lett. B\/} {\bf 166} 196--198

\bibitem{Pospelov:2007mp}
Pospelov M, Ritz A and Voloshin M~B 2008 {\em Phys. Lett. B\/} {\bf 662} 53--61

\bibitem{FAYET1990743}
Fayet P 1990 {\em Nuclear Physics B\/} {\bf 347} 743--768 ISSN 0550-3213
  \urlprefix\url{https://www.sciencedirect.com/science/article/pii/055032139090381M}

\bibitem{PhysRevLett.113.201801}
Lees J~P,  {\em et~al.\/} (BaBar) 2014 {\em Phys. Rev. Lett.\/} {\bf 113}(20)
  201801

\bibitem{PhysRevLett.119.131804}
Lees J~P,  {\em et~al.\/} (BaBar) 2017 {\em Phys. Rev. Lett.\/} {\bf 119}(13)
  131804

\bibitem{Jadach:1995nk}
Jadach S, Placzek W and Ward B~F~L 1997 {\em Phys. Lett. B\/} {\bf 390}
  298--308 (\textit{Preprint} \eprint{hep-ph/9608412})

\bibitem{Karlen:1987vk}
Karlen D 1987 {\em Nucl. Phys. B\/} {\bf 289} 23--35

\bibitem{Iwasaki:2011za}
Iwasaki Y {\em et~al.\/} 2011 {\em IEEE Trans. Nucl. Sci.\/} {\bf 58}
  1807--1815

\bibitem{NA64:2019imj}
Banerjee D {\em et~al.\/} 2019 {\em Phys. Rev. Lett.\/} {\bf 123} 121801
  (\textit{Preprint} \eprint{1906.00176})

\bibitem{Endo:2012hp}
Endo M, Hamaguchi K and Mishima G 2012 {\em Phys. Rev. D\/} {\bf 86} 095029
  (\textit{Preprint} \eprint{1209.2558})

\bibitem{Babusci:2014sta}
Babusci D {\em et~al.\/} (KLOE-2) 2014 {\em Phys. Lett. B\/} {\bf 736} 459--464
  (\textit{Preprint} \eprint{1404.7772})

\bibitem{Batley:2015lha}
Batley J~R {\em et~al.\/} (NA48/2) 2015 {\em Phys. Lett. B\/} {\bf 746}
  178--185 (\textit{Preprint} \eprint{1504.00607})

\bibitem{Andreas:2012mt}
Andreas S, Niebuhr C and Ringwald A 2012 {\em Phys. Rev. D\/} {\bf 86} 095019
  (\textit{Preprint} \eprint{1209.6083})

\bibitem{Blumlein:2011mv}
Blumlein J and Brunner J 2011 {\em Phys. Lett. B\/} {\bf 701} 155--159
  (\textit{Preprint} \eprint{1104.2747})

\bibitem{TheBelle:2015mwa}
Jaegle I (Belle) 2015 {\em Phys. Rev. Lett.\/} {\bf 114} 211801

\bibitem{Lees:2012ra}
Lees J {\em et~al.\/} (BaBar) 2012 {\em Phys. Rev. Lett.\/} {\bf 108} 211801

\bibitem{Babusci:2015zda}
Anastasi A {\em et~al.\/} (KLOE-2) 2015 {\em Phys. Lett. B\/} {\bf 747}
  365--372

\end{thebibliography}


\end{document}